\documentclass[a4paper,10pt]{article}
\usepackage{
times,
amsmath,amssymb,epsfig,graphics,graphicx}

\newcommand{\be}{\begin{eqnarray}}
\newcommand{\ee}{\end{eqnarray}}

\begin{document}

\begin{center}
{\Large\bf On the Exact Solubility in Momentum Space of
 the Trigonometric Rosen-Morse Potential  }
\end{center}
\vspace{0.02cm}

\begin{center}
C.\ B.\ Compean and M.\ Kirchbach 
\end{center}

\vspace{0.01cm}
\begin{center}
{\it Instituto de F\'{\i}sica}, \\
         {\it Universidad Aut\'onoma de San Luis Potos\'{\i},}\\
         {\it Av. Manuel Nava 6, San Luis Potos\'{\i}, S.L.P. 78290, M\'exico}
\end{center}

\vspace{0.01cm}

\begin{quote}{\bf Abstract:}
The Schr\"odinger equation with the trigonometric 
Rosen-Morse potential in flat three dimensional Euclidean space, $E_3$, 
and its exact solutions
are shown to be also exactly Fourier transformable to
momentum space, though the resulting
equation is purely algebraic and can not be cast into the canonical form
of an integral Lippmann-Schwinger equation. This is because the
cotangent function does not allow for an exact Fourier 
transform in $E_3$. In addition we recall, that the above potential
can be also viewed as an angular function of the second polar angle 
parametrizing  the three dimensional spherical surface, $S^3$,
of a constant radius, in which case the
cotangent function would allow for an exact integral transform to 
momentum space. On that basis, we obtain
a momentum space Lippmann-Schwinger type equation,  
though the corresponding wavefunctions have to be obtained numerically.

\end{quote}

\vspace{0.3cm}

\begin{flushleft}
PACS:\hspace{0.1cm} 03.65.Ge, 02.30.Uu
\end{flushleft}

\section{Introduction}
Constructing the phase spaces of quantum mechanical systems is one of the key
issues in their description. The knowledge of the potential and the
wave functions in momentum space is indispensable both in relativistic and 
nonlinear dynamics frameworks.
In particular, the trigonometric Rosen-Morse potential (Rosen-Morse I),
being of finite range, is becoming of interest in the design of 
confinement phenomena ranging form electrons in quantum dots \cite{qtm_dots}
to quarks in hadrons \cite{KC_10}, a reason for which studying its
momentum space representation appears timely.\\

\noindent
In a flat three dimensional (3D)
Euclidean position space,$E_3$, the Schr\"odinger equation,
\begin{equation}
-\frac{\hbar^2}{2\mu }\nabla^2 \psi ({\mathbf r})+
V({\mathbf r})\psi ({\mathbf r}) =E\psi({\mathbf r}),
\label{Schr_pos}
\end{equation}
provides a suitable tool for the description of systems interacting via
central potentials, many of which result exactly solvable \cite{Khare}.
The momentum space counterpart to eq.~(\ref{Schr_pos}) is obtained by means of
a Fourier transform \cite{Tang} as

\begin{equation}
\frac{{\mathbf p}^2}{2\mu }\phi ({\mathbf p}) +
\int e^{\frac{i}{\hbar }{\mathbf p }\cdot {\mathbf r}} 
V({\bf r})\psi ({\mathbf r})
d^3{\mathbf r}=E\phi({\mathbf p}),
\label{Schr_mom}
\end{equation} 
with $\phi ({\mathbf p})$ being the Fourier transform of 
$\psi ({\mathbf r})$, 
\begin{equation}
\phi ({\mathbf p})=\frac{1}{(2\pi)^{\frac{3}{2}}}
\int e^{\frac{i}{\hbar }{\mathbf p}\cdot {\mathbf r }}
\psi ({\mathbf r})d^3{\mathbf r},
\label{wafu_mom}
\end{equation}
and supposed to exist.
In cases where the Fourier transform of the potential too exists, i.e.
\begin{equation}
\frac{1}{(2\pi)^{\frac{3}{2}}}
\int e^{\frac{i}{\hbar } {\mathbf q}
\cdot {\mathbf r}} 
V({\bf r})d^3{\mathbf r} =V({\mathbf q}),
\label{IT_pot}
\end{equation}
the momentum space Schr\"odinger equation can be cast into the 
canonical form, 
\begin{equation}
\frac{{\mathbf p}^2}{2\mu }\phi ({\mathbf p}) +
\int V({\mathbf q})\phi ({\mathbf p}^\prime )
d^3{\mathbf p}^\prime=E\phi ({\mathbf p}),\quad 
{\mathbf q}={\mathbf p}-{\mathbf p}^\prime.
\label{MS_stand}
\end{equation}
The latter equation is frequently referred to as Lippmann-Schwinger equation.
{}From now onwards we introduce $F({\mathbf p})$ as a
special symbol for the integral appearing in eq.~(\ref{Schr_mom}), i.e.
we define
\begin{eqnarray}
F({\mathbf  p}) &=&
\frac{1}{(2\pi )^{\frac{3}{2}}}\int e^{\frac{i}{\hbar }{\mathbf p }
\cdot {\mathbf r}} 
V({\bf r})\psi ({\mathbf r})
d^3{\mathbf r}.
\label{vers1}
\end{eqnarray}
Provided, the integral in eq.~(\ref{IT_pot}) exists, 
an equivalent representation of eq.~(\ref{vers1})  is obtained as
\begin{eqnarray}
F({\mathbf p})&=& \frac{1}{(2\pi )^{\frac{3}{2}}}
\int V({\mathbf q})\phi ({\mathbf p}^\prime )
d^3{\mathbf p}^\prime \ .\label{Fct}
\end{eqnarray}
In case the integrals in eqs.(\ref{wafu_mom}) and ~(\ref{vers1}) 
can be taken in close forms, 
while the  integral  in eq.~(\ref{IT_pot}) does not exist, 
eq.~(\ref{Schr_mom}) is  purely algebraical. 
It is the goal of the present study to show that 
this precisely is the case of the 
trigonometrical Rosen-Morse potential (Rosen-Morse I) when considered as a 
central potential in $E_3$. The resulting momentum space is then conjugate 
to the infinite flat position space.

Various phase spaces can be designed
in considering the argument of Rosen-Morse I as
the arc of $S^3$, the minimal geodesic distance on a three dimensional
space of constant positive curvature. In particular we 
consider the arc in its projection
on the equatorial disk, a 3D space of finite volume,
and show that for this very case
the momentum space Schr\"odinger equation is no longer algebraic but
takes a form similar to the integral equation (\ref{MS_stand}).
 
The paper is organized as follows. In the next section we present
the cotangent interaction from the perspective of potential theory.
Section 3 contains the algebraic momentum space equation.
The integral momentum space equation is given in Section 4.
The paper closes with brief conclusions.

\section{The cotangent interaction from the perspective of potential theory}
The trigonometric Rosen-Morse potential  
(up to additive constants) is given by
\begin{equation}
V (\chi  )= -2B\cot\chi +\frac{\hbar^2}{2\mu d^2 } l(l+1)
\csc^2 \chi, 
\label{RMI} 
\end{equation}
where $\chi $ is a dimensionless angular variable, $\mu$ has the 
dimensionality of mass, and $d$ is a 
matching length parameter. It has the virtue to be
one among  the few potentials that can be introduced along the 
line of potential theory \cite{PotTeo}.
According to potential theory, 
functions which solve the Laplace-Beltrami
equation on a manifold of interest, the so called {\it harmonic\/}
functions,  have the remarkable
property, if employed as interactions, to preserve the symmetry of the
free geodesic motion \cite{PotTeo}.
It has been observed long ago by Schr\"odinger in ~\cite{Schr40} 
that $\cot \chi $ is a harmonic function on the 
three-dimensional (3D) surface of constant
positive curvature, the hypersphere $S^3$ embedded in a flat Euclidean space
of four dimensions, $E_4$, where $\chi$ can be viewed as
the second polar angle.
The isometry group of $S^3$ is $SO(4)$ and the cotangent function,
being harmonic there, respects this very symmetry \cite{Higgs}.
The latter is best understood by first noting  that the 
Schr\"odinger equation with the
$\csc^2$ potential is closely related to  the
eigenvalue problem of the 4D angular momentum on $S^3$.
Through the paper we consider ordinary Euclidean flat space, $E_3$,   
embedded in a 4D Euclidean space, $E_4$, and parametrize   
the 3D spherical surface $S^3$  as $x_4^2+{\mathbf r}^2=R^2$ with 
$x_4=R\cos \chi$, and $|{\mathbf r}|=R\sin\chi$. Here,
$\chi \in [0,\pi ]$ is the second polar angle in $E_4$,
 while  $R$ stands for the 
constant hyper-radius of $S^3$.
The 4D Laplace-Beltrami operator, 
${ \underline{{\overline{\vert\,\, \vert}}}},$ 
is proportional to
the operator of the squared 4D  angular momentum,
${\mathcal K}^2$, for constant radius,  and is given by
\begin{eqnarray}
{ \underline{{\overline{\vert\,\, \vert}}}} 
=-\frac{1}{R^2}{\mathcal K}^2 .
\label{lpls_4}
\end{eqnarray}
The analogue on the 2D sphere, $S^2$, of constant radius
$|{\mathbf r}|=a$,  is the well known relation
${\vec \nabla}^2 =-\frac{1}{a^2}{\mathbf L}^2 $.
Consequently,  
the  Schr\"odinger  equation for free geodesic motion on $S^3$ becomes
\begin{eqnarray}
{\Big[}\frac{\hbar^2}{2\mu R^2 } {\mathcal K}^2 
&-& {\mathcal E}{\Big]} \psi  (\chi )=0.
\label{chi_eq}
\end{eqnarray}
The ${\mathcal K}^2$ eigenvalue-problem reads
\cite{Kim_Noz}
\begin{equation}
{\mathcal K}^2 \vert K l m \rangle = K(K+2)
\vert K l m \rangle, \quad 
\vert Klm\rangle \in \left( \frac{K}{2},\frac{K}{2} \right),
\label{Casimir_O4}
\end{equation}
and the  $|Klm>$-levels belong to irreducible  
$SO(4)$ representations of the type  
\begin{equation}
|Klm\rangle \in \left(\frac{K}{2},\frac{K}{2} \right).
\label{K_levels}
\end{equation}
The quantum numbers, 
$K$, $l$, and $m$ define the eigenvalues of the respective  
four--, three-- and two--dimensional angular momentum operators
upon the  state. These quantum numbers 
correspond to the 
\begin{equation}
\begin{array}{ccccc}
SO(4)&\supset& SO(3)&\supset& SO(2),\\
K& &l&&m,
\end{array}
\label{chain}
\end{equation}
 reduction chain and
satisfy the branching rules,
\begin{equation}
K=0,1,2...,\infty , \quad l=0,1,2,.. K, \quad  m=-l,..., +l.
\label{branchrls}
\end{equation}
Therefore, the spectrum of eq.~(\ref{chi_eq}) is given by,
\begin{equation}
{\mathcal E}=\frac{\hbar^2}{2\mu R^2}K(K+2)=\frac{\hbar^2}{2\mu R^2}(K+1)^2
-\frac{\hbar^2}{2\mu R^2}.
\label{rgd_rot}
\end{equation}
It is no more but the spectrum of the 4D rigid rotor.
In terms of $\chi $ eq.~(\ref{chi_eq}) takes the explicit form
\begin{eqnarray}
\frac{\hbar^2}{2\mu R^2}\left[\frac{1}{\sin^2\chi }
\frac{\partial }{\partial \chi}
\sin^2\chi \frac{\partial }{\partial \chi } -
\frac{{\mathbf  L}^2 (\theta ,\varphi ) }{\sin^2 \chi }\right]\psi (\chi )-
{\mathcal E}\psi (\chi)&=&0.
\label{chi-chi}
\end{eqnarray}
Multiplying eq.~(\ref{chi-chi}) by $(-\sin^2\chi )$ and changing the
variable $\psi  (\chi  )$ to $\sin \chi {\mathcal S} (\chi )$,
results in the following Schr\"odinger equation
\begin{eqnarray}
\left[ 
-\frac{\hbar^2}{2\mu R^2 }\frac{\mbox{d}^2 }{\mbox{d}\chi ^2}
+U_l(\chi ) \right]
{\mathcal S} (\chi ) 
= {\mathcal E} {\mathcal S} (\chi  ),&&\nonumber\\
U_l(\chi)= \frac{\hbar^2}{2\mu R^2} l(l+1)\csc^2\chi,&&
\label{chi_free}
\end{eqnarray} 
with  $U_l(\chi, \kappa)$ now having the meaning of
 centrifugal barrier on $S^3$.
As a different  reading to eqs.~({\ref{chi_eq}), and (\ref{chi_free})
one can say that
the  $\csc^2$ potential, in representing the
centrifugal barrier on the 3D hypersphere,  has $SO(4)$ as
potential algebra.
The solutions to eq.~(\ref{chi_free}) are determined by the Gegenbauer
polynomials, $C_n^m(\cos \chi )$, as
\begin{equation}
{\mathcal S}(\chi )={\mathcal N}_{Kl}\sin^l\chi C_{K-l}^{l+1}(\cos \chi ),
\label{sol_free}
\end{equation}
where ${\mathcal N}_{Kl}$ is a  normalization constant.

Introducing now the $S^3$ harmonic function (-$2G/R \cot \chi $) as an
interaction into  eq.~(\ref{chi-chi}) results in:
\begin{eqnarray}
S^3:\quad \frac{\hbar^2}{2\mu R^2}\left[\frac{1}{\sin^2\chi }
\frac{\partial }{\partial \chi}
\sin^2\chi \frac{\partial }{\partial \chi } -
\frac{{\mathbf  L}^2 (\theta ,\varphi ) }{\sin^2 \chi }\right]\Psi (\chi )-
2G\frac{\cot\chi}{R}\Psi (\chi )-
{ E}\Psi (\chi)&=&0.
\label{chi-chi-chi}
\end{eqnarray}
The extension of eq.~(\ref{chi_free}) by same interaction amounts to
 \begin{eqnarray}
S^3:\quad \left[ 
-\frac{\hbar^2}{2\mu R^2}\frac{\mbox{d}^2 }{\mbox{d}\chi ^2}
+ \frac{\hbar^2}{2\mu R^2} l(l+1)\csc^2\chi -2G\frac{\cot\chi}{R} \right]
\psi  (\chi ) 
= E  \psi  (\chi  ), \quad B=\frac{G}{R},
\label{chi_int}
\end{eqnarray} 
and describes  $ SO(4)$ symmetric one-particles motion on $S^3$ within 
a cotangent field of a source placed on the ``Northern pole'' of
the hypersphere.
This equation is exactly solvable and the solutions are either expressed
in terms of Jacobi polynomials of a purely imaginary argument and
complex parameters that are conjugate to each-other \cite{Levai}, \cite{Dutt}
or, alternatively, in terms of the framework of the real Romanovski 
polynomials, recently elaborated in ~\cite{CK_06}, \cite{CK_07} and 
reviewed in \cite{raposo}. The spectrum of eq.~(\ref{chi_int}) now reads
\begin{equation}
E=\frac{\hbar^2}{2\mu R^2} K(K+2) -\frac{G^22\mu}{\hbar^2 (K+1)^2}.
\label{energy}
\end{equation} 
From the latter formula one reads off that
in the limit of an infinite radius, i.e. $\lim R\to \infty$,
corresponding to the expansion of the hypersphere towards the $E_3$  space,
in which the contribution of the 
curved centrifugal barrier tends to zero,
the spectrum becomes hydrogen-like \cite{Barut}. 
This is the virtue of having  introduced 
the cotangent potential in eq.~(\ref{chi_int})
as inversely  proportional to $R$.
Note, however,  that a complete $E_4$
wave equation with an angular potential introduced in this manner 
wouldn't be separable in polar coordinates.

\begin{quote}
It is furthermore important to notice that the trigonometric Rosen-Morse 
potential describes a system put on a finite volume and thereby
a confinement phenomenon, as visible through the fact that its 
spectrum is exclusively discrete. Yet, in the limit of a vanishing curvature,
$\lim \kappa \to 0$, with $\kappa =1/R^2$, the volume increases toward
infinity and the space flattens. In the more specific limit of
\begin{equation}
\lim \frac{(K+1)}{R}\stackrel{R\to \infty ,K\to \infty  }{\longrightarrow }k,
\quad k=\mbox{const}: \quad \lim E\to \frac{\hbar ^2k^2}{2\mu },
\label{sct_st}
\end{equation}
the  energies of the high-lying bound
excitations start approaching the energies of the scattering continuum
states of
the Coulomb potential \cite{Barut}. In effect, the process of
shutting down the curvature of Rosen-Morse I acquires 
features of a deconfinement transition.  
This particular property of Rosen-Morse I makes it attractive for 
the description of
confinement phenomena such as quantum dots \cite{qtm_dots} and 
quark confinement \cite{KC_10}.
 \end{quote}

Back to the 3D curved manifold under consideration,
we recall that the $\chi$ angle is measured in terms of
the  arc, $\stackrel{\smallfrown}{r}$, along the geodesic distance on $S^3$,
as
\begin{equation}
S^3:\qquad \chi =\frac{\stackrel{\smallfrown}{r}}{R}.
\label{arc}
\end{equation} 
Flat three dimensional position spaces can be approached
through various parameterizations of $\chi$.
In one of the possibilities,
the $\chi$ angle can be parametrized in terms of
the radial distance, $r$,  from origin  
within the finite volume 3D  flat space of the
 equatorial disk, ${\mathcal D}_3$, of 
the hypersphere, according to
\begin{equation}
{\mathcal D}_3:\quad 
\chi =\sin^{-1}\left( \frac{r}{R}\right), \quad r\in [0,R],
\quad {\mathbf r}\in
{\mathcal D}_3\subset S^3.
\label{RW}
\end{equation}
 This is the parametrization used in the Robertson-Walker metric of a
closed universe for which eq.~(\ref{chi-chi-chi}) takes the form
\begin{eqnarray}
{\mathcal D}_3:\quad 
{\Big[}-\frac{\hbar^2}{2\mu R^2}\frac{
\sqrt{1-\frac{r^2}{R^2}}}{r^2}
\frac{\rm d}{{\rm d}r}\sqrt{1-\frac{r^2}{R^2}}r^2
\frac{\rm d}{{\rm d}r}
+\frac{\hbar^2}{2\mu}\frac{l(l+1)}{r^2}  -2G
\frac{
\sqrt{1-\frac{r^2}{R^2}}
}{r}{\Big]}
\psi \left(\frac{r}{R}\right)&=&E \psi \left(\frac{r}{R}\right), \nonumber\\
 \psi \left(\frac{r}{R}\right):\stackrel{\mbox{df}}{=}
\Psi \left(\sin^{-1}\left(\frac{r}{R}\right)\right).
\label{QDT}
\end{eqnarray}
In the physics of quantum dots mentioned above, another
parametrization, $\chi =\tan^{-1}r/R $,  has been used \cite{qtm_dots}
which projects the surface of the hypersphere on the infinite 3D flat 
plane tangential to the North pole. Also in this case, 
the position space Schr\"odinger equation
contains beyond the standard kinetic piece, several gradient terms 
 needed to recover the finite range confinement, as also 
happens in  eq.~(\ref{QDT}).

Instead, the parametrization used by Schr\"odinger \cite{Schr40} himself
and subsequently adopted by  SUSYQM \cite{Khare} is,
\begin{equation}
E_3:\quad \chi =\frac{r}{R}\pi=\frac{r}{d}, \quad d=\frac{R}{\pi},
\quad \frac{r}{R}\in [0,1].\label{flat}
\end{equation} 
In the latter parametrization  
the $(\cot+\csc^2)$ interaction formally acquires features of 
a central $E_3$  flat space potential and the respective 
Schr\"odinger equation obtained now from eq.~(\ref{chi_int})
takes the well known standard form of wide spread,
\begin{equation}
E_3:\quad \left[ 
-\frac{\hbar^2}{2\mu d^2}\frac{\mbox{d}^2 }{\mbox{d} r^2}
+ \frac{1}{d^2} \frac{\hbar^2}{2\mu } l(l+1)\csc^2\left( \frac{r}{d}\right) -
2B\cot\left( \frac{r}{d}\right) \right]
\psi  \left(\frac{r}{d}  \right) 
= E \psi  \left(\frac{r}{d}  \right), \quad B=\frac{G}{d}.
\label{RM_SQM}
\end{equation}
It is the goal of the present study to transform 
above eqs.~(\ref{RM_SQM}) and (\ref{QDT}) to momentum space. 
We hope that the reader will not be confused by the repeated use of 
same symbol $\psi$ as a wavefunction but will figure out
it  significance from the relevant context.
Before moving further we wish to emphasize that the variable changes
in the master equation (\ref{chi_int}) of course do not affect
physics and the energy spectra of 
eqs.~(\ref{chi_int}), (\ref{RM_SQM}), and (\ref{QDT}) stay same.

\section{The algebraic momentum space equation}
Our first point is that the exact solutions of eq.~(\ref{RM_SQM})
constructed earlier by us in ~\cite{CK_06}, \cite{CK_07}
allow for exact Fourier transforms to momentum space.
As a reminder, as well as for the sake of self-sufficiency  of the 
presentation 
and illustrative purposes, we
here bring these solutions for the first three $S$ wave states, 
displayed in Fig.~1.
\begin{figure}
\resizebox{0.80\textwidth}{7.5cm}
{\includegraphics{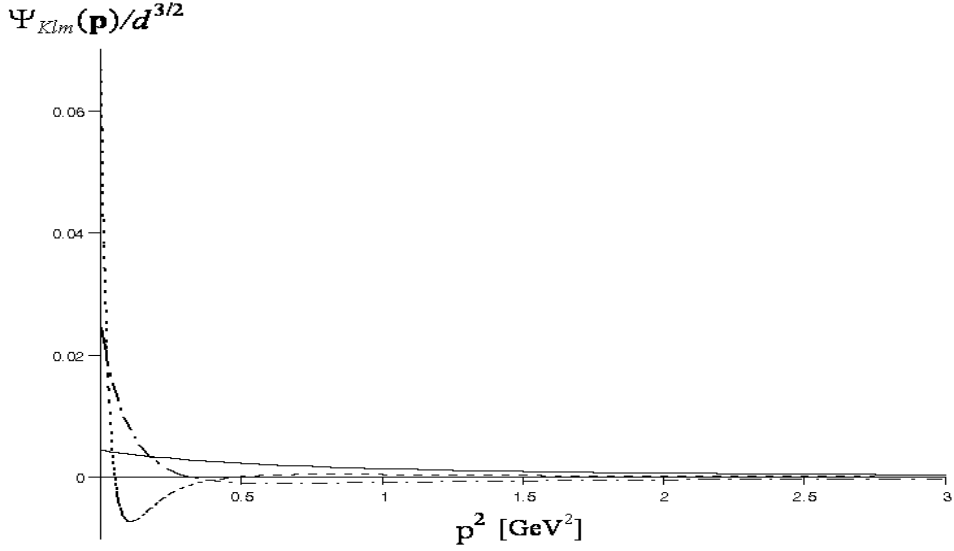}}
\caption{Momentum dependence of the lowest $S$ wave states 
in eq.~(\ref{SW_3}). The solid, dashed-dotted, and dotted 
lines correspond to $K=0$, $K=1$, and $K=2$, respectively.
\label{fiu1}}
\end{figure}

Subsequently, the solutions of eq.~(\ref{RM_SQM}) will be specified by
the quantum numbers defined in eqs.~(\ref{K_levels}), (\ref{chain}), 
and (\ref{branchrls}) as $\Psi_{Klm}(z)$ with $z=r/d$.  
The following expressions hold valid:
\begin{eqnarray}
\Psi_{0,0,0}(z) & = & N_0\frac{\sin z}{z} 
e^{-b z}\ , \quad z=\frac{r}{d}, \quad b=\frac{B2 \mu d^2}{\hbar^2},\\
\Psi_{1,0,0}(z) & = & N_1 \left(\cos z - \frac{b}{2}\sin z\right) 
\frac{\sin z}{z} e^{-b \frac{z}{2}}\ ,\\
\Psi_{2,0,0}(z) & = & N_2 \left(3 \cos^2 z - 2 b\sin z \cos z + 
\left(2\left(\frac{b}{3}\right)^2-1\right)\sin^2 z\right) \frac{\sin z}{z} 
e^{-b \frac{z}{3}}\ .
\label{SW_1}
\end{eqnarray}
The normalization constants is
given by 
\begin{equation}
N_K = (-1)^K \left(\frac{1}{\pi d^{3}}\right)^{1/2} 
\left(\frac{b}{K+1}\left(\left(\frac{b}{K+1}\right)^2+
(K+1)^2\right)\right)^{1/2}\ ,
\end{equation}
for $K=0,1,2,...$.
In introducing the short-hand ${\bar p}$ as
\begin{equation}
{\bar p}=\frac{pd}{\hbar}, \quad p=|{\mathbf p}|,
\end{equation}
the following  Fourier transforms  of the above wavefunctions 
are now obtained in close form:
\begin{eqnarray}
\phi_{0,0,0}(\bar p) & =&  
\left(\frac{2bd}{\hbar}\right)^{\frac{3}{2}}\frac{(b^2+1)^{\frac{1}{2}}}
{\pi}\frac{1}{{\bar p}^4+2(b^2-1){\bar p}^2+2b^4+1}\ ,\\
\phi_{1,0,0}(\bar p)  &=&  -
\frac{2^{\frac{1}{2}}}{\pi}
\left(\frac{2bd}{\hbar}\right)^{\frac{3}{2}}
\left(\left(\frac{b}{2}\right)^2+1\right)^{\frac{1}{2}}\nonumber\\
&&\frac{{\bar p}^2-\left(\frac{b}{2}\right)^2-2}{{\bar p}^6+
\left(3\left(\frac{b}{2}\right)^2-8\right){\bar p}^4 + 
\left(3\left(\frac{b}{2}\right)^4+16\right){\bar p}^2 + 
\left(\left(\frac{b}{2}\right)^4+8\left(\frac{b}{2}\right)^2+16\right)
\left(\frac{b}{2}\right)^2}\ ,
\end{eqnarray}
\begin{eqnarray}
\phi_{2,0,0} (\bar p)  =  \frac{
\sqrt{3}}{\pi}\left( \frac{2bd}{\hbar}\right)^{\frac{3}{2}}
\left( \left( \frac{b}{3}\right)^2 +1\right)^{\frac{1}{2}}
\left(
{\bar p}^4 -\left( 10\left( \frac{b}{3}\right)^2 +22\right)\frac{{\bar p}^2}{3}
+\left( \frac{b}{3}\right)^4
+\frac{22}{3}\left( \frac{b}{3}\right)^2
+\frac{19}{3}
\right)&&
\nonumber\\
{\Big[}{\bar p}^8 +4\left( \left( \frac{b}{3}\right)^2-5\right){\bar p}^6
+2\left( 
\left( \frac{b}{3}\right)^4
-10\left(\frac{b}{3} \right)^2
+59\right){\bar p}^4
+4\left( 
\left( \frac{b}{3}\right)^6
+5\left(\frac{b}{3} \right)^4
+23\left( \frac{b}{3}\right)^2 +45
\right){\bar p}^2&&\nonumber\\
+\left(
\left( \frac{b}{3}\right)^8
+20\left( \frac{b}{3}\right)^6
+118\left( \frac{b}{3}\right)^4
+20\left( \frac{b}{3}\right) 
+81
\right)
{\Big]}^{-1}&&
\label{SW_3}
\end{eqnarray}
In employing  symbolic programs, the 
existence of exact Fourier
transforms  for the other states can be checked. 
The same applies to the related integrals  in eq.~(\ref{vers1}).
As a consequence, the following algebraic equation emerges:

\begin{eqnarray}
\left(E-\frac{{\mathbf p}^2}{2\mu }\right)\phi_{Klm}({\mathbf p})&=&
F_{Klm}({\mathbf p}),\nonumber\\
F_{Klm}({\mathbf p})&=&
\frac{1}{(2\pi )^{\frac{3}{2}}}\int e^{\frac{i}{\hbar }{\mathbf p }
\cdot {\mathbf r}} 
\left( -\frac{2G}{d}\right)\cot \left( \frac{r}{d}\right)
\Psi_{Klm} \left(\frac{r}{d}\right)Y_l^m(\theta ,\varphi)
{\mathrm d}^3{\mathbf r}.
\label{algbr_momsp}
\end{eqnarray}
Below we bring the explicit expressions for $F_{Klm}$ 
for the two lowest $S$ states,
displayed in Fig.~2, for illustrative purposes,
\begin{eqnarray}
F_{0,0,0}({\bar p})&=&
\frac{1}{\pi}
\left(\frac{2bd}{\hbar}\right)^{\frac{3}{2}}(b^2+1)^{\frac{1}{2}}
\frac{1-b^2-{\bar p}^2}{{\bar p}^4+2(b^2-1){\bar p}^2+2b^4+1}\\
F_{1,0,0}({\bar p})&=&
-\frac{2^{\frac{1}{2}}}{\pi}
\left(\frac{2bd}{\hbar}\right)^{\frac{3}{2}}\left(
\left(\frac{b}{2}\right)^2+1\right)^{
\frac{1}{2}}\nonumber\\
&&\frac{
\left(\frac{b}{2}\right)^4 -2\left(\frac{b}{2}\right)^2-8
+6{\bar p}^2 -{\bar p}^4
}
{{\bar p}^6+\left(3\left(\frac{b}{2}\right)^2-8\right){\bar p}^4 + 
\left(3\left(\frac{b}{2}\right)^4+16\right){\bar p}^2 + 
\left(\left(\frac{b}{2}\right)^4+8\left(\frac{b}{2}\right)^2+
16\right)\left(\frac{b}{2}\right)^2}\ .
\end{eqnarray}

\begin{figure}
\resizebox{0.80\textwidth}{7.5cm}
{\includegraphics{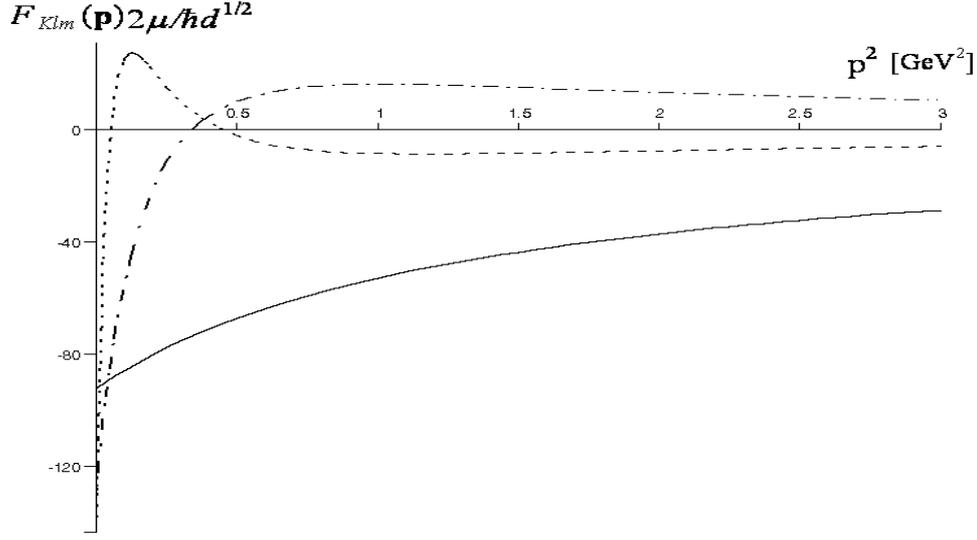}}
\caption{Momentum dependence of the $F_{Klm}$ quantity in 
eq.~(\ref{algbr_momsp})
for the lowest $S$ wave states. The solid, dashed-dotted, and dotted lines
refer to $K=0$, $K=1$, and $K=2$, respectively. 
\label{fiu2}}
\end{figure}

\section{The integral momentum space equation}
Equation (\ref{algbr_momsp}) from the previous section
is quite transparent indeed, however it doesn't make the
momentum space potential explicit. This may be perceived as a 
shortcoming by T-matrix based spectroscopic approaches \cite{T-matrix}. 
Below we shall show that
such a shortcoming is  not suffered by the 
Fourier transformed eq.~(\ref{QDT}), obtained from the
master equation (\ref{chi_int}) via the change of variable according to
eq.~(\ref{RW}).
In introducing the short hand,
\begin{equation}
x=\frac{r}{R}, \quad r=|{\mathbf r}|,
\label{x-def}
\end{equation}
eq.~(\ref{QDT}) equivalently rewrites to
\begin{equation}
-\frac{\sqrt{1-x^2}}{x^2}\frac{\mathrm d}{{\mathrm d}x}
x^2\sqrt{1-x^2}\frac{{\mathrm d}}{{\mathrm d} x}\psi(x)
+\frac{l(l+1)}{x^2}\psi (x)
-2b\frac{\sqrt{1-x^2}}{x}\psi (x) -\epsilon \psi (x)=0.
\label{QDT_1}
\end{equation}
where we in addition have divided by $\hbar^2/2\mu R^2$ and switched to
dimensionless constants by introducing
the two new notations,
\begin{equation}
b=\frac{G}{R}\left( \frac{\hbar^2}{2\mu R^2}\right)^{-1},\quad
\epsilon =E\left( \frac{\hbar^2}{2\mu R^2}\right)^{-1}.
\label{constants}
\end{equation}
After some algebraic manipulations equation~(\ref{QDT_1}) takes the form
\begin{equation}
-\left( 
\nabla^2
-
\frac{1}{x^2}
\frac{{\mathrm d}}{{\mathrm d}x}x^2
\frac{{\mathrm d}}{{\mathrm d}x}x^2 +
3x \frac{{\mathrm d}}{{\mathrm d}x}  +6
\right)\psi (x)
-2b\frac{\sqrt{1-x^2}}{x}\psi (x) -\epsilon \psi(x)=0.
\label{QDT_2}
\end{equation}
This is the equation which we shall Fourier transform to momentum space.
We first obtain the integral transform of the interaction piece of
the Hamiltonian as
\begin{eqnarray}
\frac{-2b}{(2\pi )^{\frac{3}{2}}}\int
{\mathrm d}^3{\mathbf x} e^{
i{\mathbf k}\cdot {\mathbf x }   }
\frac{\sqrt{1-x^2}}{x}\psi (x)&=&
-\frac{1}{2(2\pi )^{\frac{3}{2}}}\int
{\mathrm d}^3{\mathbf k}^\prime \int {\mathrm d}^3{\mathbf k}^{\prime\prime}
\Pi (|{\mathbf q}|)
\frac{\left(J_0({\mathbf k}^{\prime \prime })-
2J_1(
{\mathbf k}^{\prime\prime})\right)}{
\left(
{\mathbf  k}^{\prime \prime}\right)^2 
}\phi ({\mathbf k}^\prime ), \nonumber\\
{\mathbf q}&=&{\mathbf k}-{\mathbf k}^\prime -{\mathbf k}^{\prime\prime}.
\label{QDT3}
\end{eqnarray}
Here,  use has been made of
\begin{eqnarray}
\sqrt{1-x^2}&=&-\frac{1}{(2\pi )^{\frac{3}{2}}}\int
{\mathrm d}^3{\mathbf k}^{\prime \prime }
e^{-i{\mathbf k}^{\prime \prime }\cdot {\mathbf x}}
\sqrt{\frac{\pi }{2 }}\frac{\left( J_0({\mathbf k}^{\prime \prime })-
2J_1(
{\mathbf k}^{\prime\prime})\right)}{
\left(
{\mathbf  k}^{\prime \prime}\right)^2 
}, \quad x=|{\mathbf x}|\leq 1,
\label{G_mom}
\end{eqnarray}
and
\begin{eqnarray}
\frac{1}{x}&=& \frac{1}{(2\pi )^{2}}
\int {\mathrm d}^3{\mathbf q}\, e^{
-i{\mathbf q}\cdot {\mathbf x} }
\frac{
\sin^2 \frac{|{\mathbf q}|}{2}
}
{
\left(\frac{{\mathbf q}}{2}\right)^2
}=
\frac{1}{(2\pi )^{2}}
\int {\mathrm d}^3{\mathbf q}\, e^{
-i{\mathbf q}\cdot {\mathbf x} }
\frac{1}{(-b)}\Pi (|{\mathbf q}|),
\quad x=|{\mathbf x}|\leq 1.
\end{eqnarray}
In the latter equation we worked in the momentum-space potential,
$\Pi (|{\mathbf q}|)$, obtained earlier by us in ref.~\cite{CK_09} as
\begin{eqnarray}
\Pi(|{\mathbf q}|)&=&(-b)\frac{\sin^2 \frac{|{\mathbf q}|}{2} }{
\left(
\frac{{\mathbf q}}{2}
\right)^2
},
\label{propagator}
\end{eqnarray}
the $E_4$ Fourier transform of $(-2b\cot \chi )$ for the elastic scattering
case, $q_4=0$,  which we calculated
using the $S^3$ integration volume according to
\begin{eqnarray}
4\pi \Pi
( |\mathbf{q}| ) &=&
-2b\int_0^\infty d|x| |x|^3 \delta (|x| -1)\int_0^{2\pi} d \varphi 
\int_0^\pi 
d\theta \sin\theta \nonumber\\
&&\int_{0/\frac{\pi}{2}}^{\frac{\pi}{2}/\pi } d \chi \sin^2\chi 
e^{i|\mathbf{ q}|\frac{\sin\chi}{\sqrt{\kappa}}|\cos\theta}
\cot \chi .
\label{b2}
\end{eqnarray}
We used the $\delta (|x| -1)$ function in order to restrict $E_4$ to 
$S^3$. In the context of the present study and the  dimensionless variable in
equation~(\ref{x-def}), $S^3$ 
presents itself as the unit
hypersphere.\\

\noindent
Finally, transforming the  pieces of the kinetic part in 
eq.~(\ref{QDT_2}) amounts to
\begin{eqnarray}
\frac{1}{
(2\pi )^{\frac{3}{2}}}\int {\mathrm d}^3{\mathbf x}
e^{i{\mathbf k}\cdot {\mathbf x}}
x\frac{{\mathrm d}\psi (x)}{{\mathrm d}x}=-3\phi ({\mathbf k}) -
{\mathbf k}\cdot \nabla _{{\mathbf k }}\phi ({\mathbf k}),
\end{eqnarray}

and

\begin{eqnarray}
\frac{1}{(2\pi )^{\frac{3}{2}}}\int
{\mathrm d}^3{\mathbf x}e^{i{\mathbf k}\cdot {\mathbf x}}
{x}
\frac{{\mathrm d}}{{\mathrm d}x}x
\frac{{\mathrm d}}{{\mathrm d}x}x\psi (x) &=&
{\mathbf k}\cdot {\nabla }_{{\mathbf k}}
{\mathbf k}\cdot {\nabla }_{{\mathbf k}}\phi ({\mathbf k})
+6\, {\mathbf k}\cdot {\nabla }_{{\mathbf k}}\phi ({\mathbf k}) 
+9\phi ({\mathbf k}),
\end{eqnarray}
respectively.

Putting it all together 
the integral momentum space equation becomes
\begin{eqnarray}
\left( {\mathbf k}^2 
+{\mathbf k}\cdot {\nabla }_{{\mathbf k}}
{\mathbf k}\cdot {\nabla }_{{\mathbf k}}
+4\, {\mathbf k}\cdot {\nabla }_{{\mathbf k}}
+3\right)\phi ({\mathbf k})-
 \frac{1}{2(2\pi )^{\frac{3}{2}}}\int
{\mathrm d}^3{\mathbf k}^\prime \phi ({\mathbf k}^\prime )
 \int {\mathrm d}^3{\mathbf k}^{\prime\prime}
\Pi (|{\mathbf q}|)
\frac{J_0({\mathbf k}^{\prime \prime })-
2J_1(
{\mathbf k}^{\prime\prime})}{
\left(
{\mathbf  k}^{\prime \prime}\right)^2 
}
&=&\epsilon \phi ({\mathbf k}).\nonumber\\
\end{eqnarray}
This equation is satisfied by the numerical Fourier transforms of the
wavefunctions from position space.

\section{Conclusions}
We recalled that the most natural view on the
trigonometric Rosen-Morse potential
appears within the context of a particle moving in
the cotangent field of a source placed at one of the poles
of the three-dimensional spherical surface, $S^3$, of a
constant radius, embedded within a four-dimensional 
Euclidean space, $E_4$. Within this context the trigonometric Rosen-Morse
potential has been considered as
an angular function of the arc  of the minimal geodesic
distance on $S^3$ and the related Schr\"odinger equation
has been presented in eq.~(\ref{chi_int}).
This general view on Rosen-Morse I
has the virtue of making its $SO(4)$ symmetry obvious. Indeed, 
within this context, the $\csc^2\chi $ part plays the part of a 
centrifugal barrier on $S^3$ whose isomerty group 
is $SO(4)$, a symmetry respected by the  $\cot \chi  $ piece,
known to be a harmonic function on this manifold. 
We then considered two different parameterizations
of the arc in eq.~(\ref{arc}) and  presented them
in eqs.~(\ref{RW}), and (\ref{flat}).
We wrote down  the related position space  Schr\"odinger equations 
in eqs.~(\ref{QDT}), and (\ref{RM_SQM}), respectively.
We noted that equation~(\ref{RM_SQM}) coincides with the Schr\"odinger equation
with Rosen-Morse I manged by SUSYQM, where the potential is treated as
central  in ordinary three-dimensional flat Euclidean space,
$E_3$. 
We subjected eqs. ~(\ref{QDT}) and (\ref{RM_SQM}) 
to ordinary 3D  Fourier transformations to momentum space. We 
showed that in the first case one finds a genuine integral, in the second
an exactly solvable algebraic  equation.

We expect our findings to be of use in  momentum space
many-body frameworks based on the T-matrix.

\vspace{0.53cm}
\noindent
Work supported by CONACyT-M\'{e}xico under grant number
CB-2006-01/61286.

\end{document}